\documentclass[twocolumn,showpacs,preprintnumbers,amsmath,amssymb,superscriptaddress]{revtex4}

\usepackage{blindtext}
\usepackage{graphicx}% Include figure files
\usepackage{dcolumn}% Align table columns on decimal point
\usepackage{bm}% bold math

\begin{document}

%\preprint{APS/}

\title{Cavity Quantum Electrodynamics with Dressed States of a Superconducting Artificial Atom  }

\author{Yu-Han Chang}
\affiliation{Department of Physics, National Chung Hsing University, Taichung 402, Taiwan}
\author{Dmytro Dubyna}
\affiliation{Department of Physics, National Chung Hsing University, Taichung 402, Taiwan}
\author{Wei-Chen Chien}
\affiliation{Department of Physics, National Chung Hsing University, Taichung 402, Taiwan}

\author{Chien-Han Chen}

\affiliation{Department of Physics, National Changhua University of Education, Changhua, Taiwan}

\author{Cen-Shawn Wu}

\affiliation{Department of Physics, National Changhua University of Education, Changhua, Taiwan}
\author{Watson Kuo}
\email{wkuo@phys.nchu.edu.tw}
\affiliation{Department of Physics, National Chung Hsing University, Taichung 402, Taiwan}

\begin{abstract}
We experimentally studied the microwave response of a transmon artificial atom coupled to two closely spaced resonant modes. When the atom is under driven with one of the modes, the atom state and mode photons are superposed, forming the dressed states. Dressed states with 1st, 2nd and 3rd excited states of the atom were prepared and probed via the strong coupling to the other resonant mode from the point of view of cavity quantum electrodynamics. The transmission of the probe tone is modulated by the driving microwave amplitude, displaying multi-photon process associated with the inter-atomic level transitions. Our system provides an easy method to study the dressed states by driving one mode and probing the Landau-Zener transition of the other.
\end{abstract}
\maketitle

%\section{Introduction}
The development of superconducting quantum circuits has led to realization of quantum optics at microwave frequencies.\cite{Nori11} In particular, an architecture that consists of microwave transmission lines and micrometer-sized superconducting circuit provides a great enhancement in coupling strength between electromagnetic waves and quantum circuit, which is also called ``artificial atom". Taking the advantage of strong coupling, a broad range of quantum optical phenomena, such as single atom spectroscopy\cite{Tsai10a}, dressed states\cite{Del07}, amplification of light\cite{Tsai10b}, lasing\cite{Tsai07}, Mollow triplets, Aulter-Townes splitting\cite{Sill09,Hoi13} and electromagnetically-induced transparency\cite{Tsai10c} have been demonstrated.
In addition, almost arbitrary tunability of superconducting quantum circuits in the GHz range contributes observation of different phenomena in artificial atoms without restrictions in operation frequency.

Strongly driven two level system(TLS) may produce so-called Landau-Zener (LZ) transition\cite{RN5871} that can be demonstrated in various systems, such as an atom in an intensive laser field. Recently, LZ transition has been investigated in superconducting quantum circuits, such as Cooper-pair boxes(CPB)\cite{Naka01, Silla06, Del07, La09} and flux qubits\cite{Oliv05, Sun09}. The phenomenon observed in these quantum circuits is significant and impressive due to the huge driving fields that can be achieved. In most cases, the TLS under a drive can be written as
$
{H}/{\hbar\omega_d}=-\varepsilon(t)\sigma_z-\lambda\sigma_x.
$
Here $\sigma_{x,z}$ are Pauli matrices, $\varepsilon(t)$ and $\lambda$ are respectively the bias and inner coupling of the two states in unit of $\hbar\omega_d$, where $\omega_d$ is the driving angular frequency. The driving field is treated classically and parameterized by $\varepsilon(t)=\varepsilon_0+\varepsilon_1\sin\omega_d t$. Alternatively, a picture of dressed state considers the drive as quantized excitations of a harmonic oscillator and yields an equivalent hamiltonian, 
\begin{equation}
\frac{H}{\hbar\omega_d}=-\varepsilon_0\sigma_z-\lambda\sigma_x+ a^\dag a+\eta(a^\dag+a)\sigma_z.
\end{equation}
$\eta$ can be interpreted as the coupling energy of the TLS and driving photon field,  $a$ and $a^\dag$ are annihilation and creation operators for the driving field quanta, $N$, which is related to the driving amplitude through $\varepsilon_1\sim\eta\sqrt{\langle N\rangle}$. Under strong drive, multiphoton processes occur when the TLS energy matches the energy of an integer number of photons.  The population as a function of driving amplitude shows oscillatory behavior, a general signature of Landau-Zener-St\"uckelberg (LZS) interferometry.\cite{Oliv05, Silla06, Sun09, La09}

\begin{figure}
%\graphicspath{{fig/}}
\includegraphics[width=0.3\textwidth, angle =270]{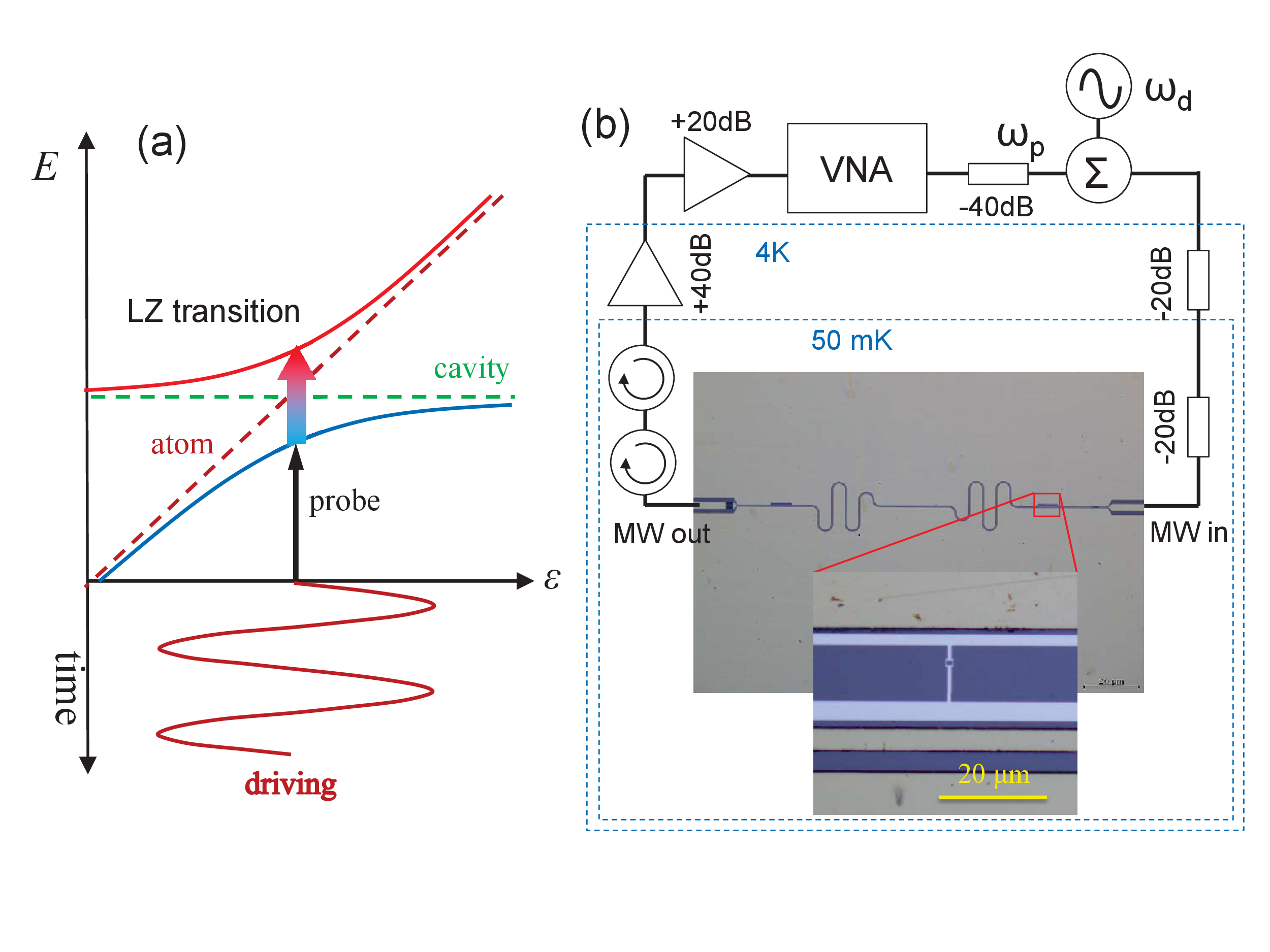}
\caption{
(a) The scheme of the Z-driving on a qubit coupled to a cavity.  The drive may produce the Laudau-Zener transition and modulate the probe tone in resonance to the lower energy branch. (b) The schematic of the 2-tone measurement. The chip under test contains a transmission type cavity. The transmon is made in the slot of the cavity. 
}\label{fig0}
\end{figure}

	The system similar to Eq.(1) can be built for a general two level atom, of which energy eigenstates are $|q=0\rangle$ and $|q=1\rangle$, is driven by the photon field. In this picture, it is described by the model 
\begin{equation}
\frac{H}{\hbar\omega_d}=-\frac{\omega_a}{2\omega_d}\sigma_z+ a^\dag a+V_\eta.
\end{equation}
$\hbar\omega_a$ is the atom energy and the interaction $V_\eta$ can be $V_\eta=\eta(a^\dag+a)\sigma_z$ or  $\eta(a^\dag+a)\sigma_x$, depending on the analogy to the application of an ac $z$ or $x$ field to the spin magnetic resonance. The ``X-driving" model is also called Rabi model and would be identical to that described in Eq.(1) by setting $\varepsilon_0=0$, $\lambda=\omega_a/2\omega_d$ and interchanging $x$ and $z$. However, the dressed states are slightly different from those in CPB(or ``Z"-driving) because they preserve the $Z_2$ symmetry in such a way that the parity of  $p=(q+N)$ is conserved under the interaction. Recently, the integrability and the route to exact solution of the model has been shown.\cite{Braak11} 

	In any case, when the dressed atom is coupled to a cavity, the interaction and cavity Hamiltonian can be described as
$
V_r/\hbar\omega_r= b^\dag b+g(b^\dag+b)\sigma_x ,
$
where $b$ and $b^\dag$ are annihilation and creation operators of the cavity photons and $g$ is the coupling strength between the atom and cavity photon. The case of ``Z-driving" in the cavity quantum electrodynamics(QED) system can be easily understood by using the classical picture depicted in Fig. 1(a). When the atom is biased in resonance to the cavity, the system shows an anti-crossing energy spectrum and the drive on bias $\varepsilon$ could lead to LZ transition, modulating the population on the lower energy branch that obeys the Bessel dependence. On the other hand, $Z_2$ symmetry in the X-driving case could be broken by the interaction $V_r$, therefore we could expect the that a similar modulation of population by the dressed photon number $N$ can be observed in this system. The LZS interferometry should be displayed when we use a probe in resonance to the lower energy branch. The recent advances in circuit cavity QED for all kinds of artificial atoms pave the way to realization of this operation without much difficulty. In addition, some theoretical works suggest that multi-photon Jayes-Cumming model is possible to be simulated by a driven atom.\cite{Pue19}

In this paper, we report the study of cavity QED with dressed states utilizing a transmon-type artificial atom in a co-planar waveguide(CPW) resonator. The transmon is strongly coupled to two different photon modes, supplying required $g$ and $\eta$ in this model. The LZ transition was controlled by the driving tone amplitude and monitored via the transmission of the probe tone. The transmission amplitude demonstrated LZS interferometry, which follows the Bessel behavior as a function of driving amplitude with multi-photon processes up to $m=3$. In other words, our system effectively exhibits a 2 or 3-photon Jayes-Cumming model depending on the selection of driving frequency and amplitude.

%\section{Experimental Methods}
As illustrated in Fig. 1(b), the sample is a transmon-type artificial atom coupled to a co-planar waveguide resonator. The artificial atom and the resonator were made of Al and Nb, respectively on a Si substrate. The transmission-type resonator has an input port with a coupling capacitance 8 fF and output port with a coupling capacitance 64 fF. The transmon is coupled to the signal line through a large capacitance 13 fF. The total capacitance of the transmon, including those from junctions and to the ground plane is estimated as $\sim$38 fF. The Josephson coupling energy $E_J/h$ is estimated as 90 GHz in zero magnetic field, which gives $E_J/E_C= 180$. A superconducting quantum interference device in the transmon allows the tuning of energy levels by a magnetic field.  The sample was cooled down in a dilution refrigerator with a base temperature below 40 mK. The transmission of the sample is measured by using a commercial vector network analyzer or a spectrum analyzer.

\begin{figure}
%\graphicspath{{fig/}}
\includegraphics[width=0.2\textwidth, angle =270]{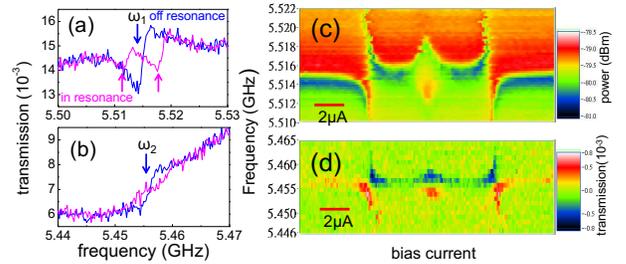}
\caption{
(a)(b) The microwave transmissions nearby the two resonances at 5.514 GHz(a) and 5.455 GHz(b).  The curves are for the transmon in resonance (red) and off-resonance(blue) with the 5.514GHz mode. (c)(d) The microwave transmission power as a function of magnet bias current and frequency in the vicinity of 5.514GHz(c) and the transmission amplitude nearby 5.455GHz(d). Both plots show the vacuum Rabi effect due to the coupling of the transmon to the resonant modes. 
}\label{fig1}
\end{figure}

\begin{figure}
%\graphicspath{{fig/}}
\includegraphics[width=0.3\textwidth, angle=270]{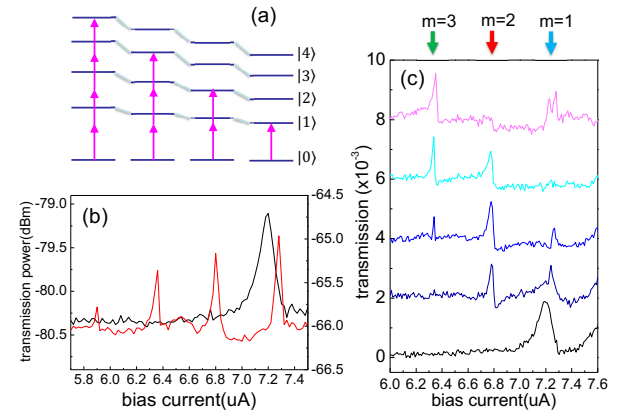}
\caption{(a) Schematic illustration of the multi-photon processes. (b)
The transmitted microwave power as a function of bias current for different probe powers at 5.5145 GHz. For the lower power $P_p=-125$ dBm, only $|0\rangle$ to $|1\rangle$ transition was observed(black). For the higher power $P_p=-110$ dBm, $|0\rangle$ to $|m\rangle$ transitions with $m=$1, 2 , 3 and 4 were observed(red).  (c) The transmission of the probe microwaves at $\omega_p/2\pi=5.513$ GHz as a function of transmon bias current for various driving microwave powers at $\omega_d/2\pi=5.455$ GHz. From the bottom to top, the driving power: off, $P_d=$-108, -104.5, -102 and -100 dBm. The higher current, the lower $E_J$ and energy level spacings. In the absence of driving microwaves, the transmission shows a resonance peak at 7.2 $\mu$A, corresponding to transition energy $\omega_{10}=\omega_p$.  For the higher driving powers, the resonance peaks appear at 6.8 $\mu$A and 6.3  $\mu$A, corresponding to $\omega_{20}=2\omega_p$ and $\omega_{30}=3\omega_p$, respectively. 
}\label{fig2}
\end{figure}

In zero magnetic field, the transmon is off-resonance with both resonant frequencies of $\omega_1/2\pi=$5.514 GHz and $\omega_2/2\pi=$5.455 GHz. Fig.  \ref{fig1} (a)(b) respectively show the microwave transmission amplitude as a function of frequency in vicinity of the two modes when the transmon is off-resonance with both modes(blue) and in-resonance with the higher mode(red). Each of the mode showed coupling with the transmon for their resonant frequencies are shifted by several MHz when the transmon and modes are in resonance. 
Fig. \ref{fig1} (c) and (d) respectively show the microwave transmission amplitude as a function of magnet bias current and frequency in the vicinity of $\omega_1$ and $\omega_2$, respectively. They both show clear anti-crossing structure when the transmon and a resonant mode is in resonance. Because the loss of the resonant mode is smaller than the coupling energy, we may resolve the splitting and find the mode-transmon coupling energies, $2g_1/h=$10 MHz, and $2g_2/h=$5 MHz. Though they are not as big as state-of-the-art values but sufficient for our operation. This system is interesting because the frequencies of two resonant modes and transmon transition can be tuned very close to each other. %In particular, when the detuning of the transmon resonance and resonance mode are larger than $2g$ but no too large, there will be frequency shift in resonance mode due to transmon quantum state. 

When the probe frequency is slightly detuned from $\omega_1/2\pi$, we may observe the transmon resonance due to the ground $|0\rangle$ to the 1st excited state $|1\rangle$ transition. This transition is shown in Fig.  \ref{fig2}(b) as a resonance peak at the bias current of 7.3 $\mu$A in the transmitted microwave power(after amplification) at the probe power $P_p=-125$ dBm and frequency of 5.5145 GHz(black curve). Due to parity symmetry,  the transition matrix element for $|0\rangle$ to $|2\rangle$ is very small for transmons. Such a selection rule for transitions associated with parity change makes the transmon a $\Xi$-type atom. Therefore, one cannot observe the second excited state  $|2\rangle$ simply by using 1-photon spectroscopy as above. Nevertheless, because the large coupling strength between the transmon and cavity, the $|0\rangle$ to $|2\rangle$ transition can occur with a 2-photon process in a large probe power. Indeed, in our sample we can observe certain multiple photon absorption process when the transition to a higher excited state is in resonance with a strong probe microwave. The red curve in Fig. \ref{fig2}(b) shows that for the high power  $P_p=-110$ dBm, three more resonance peaks develop with an equal spacing of 0.45$\mu$A for the lower currents. The higher power is applied, the more peaks develop in the vicinity of the original resonance structure. The resonance peak which develops at a higher probing power should related to a process involving more photons. For example, from the power dependence on the peak height, we can attribute the peak more close to the original one to be the 2-photon process.  Taking into account the energy shift introduced to the transmon with the bias current, we may judge that this is  a signature of two-photon transition from ground state $|0\rangle$ to the 2nd excited state $|2\rangle$. The energy level spacing  $\hbar\omega_{12}$ of transition is slightly smaller than $\hbar\omega_{01}$, the two-photon absorption will appear at a smaller bias current in comparison with the single photon absorption for $|0\rangle$ to $|1\rangle$ transition. The proposed scheme of all multi-photon processes is illustrated in Fig. 3(a).

\begin{figure}
%\graphicspath{{fig/}}
\includegraphics[width=0.16\textwidth, angle=270]{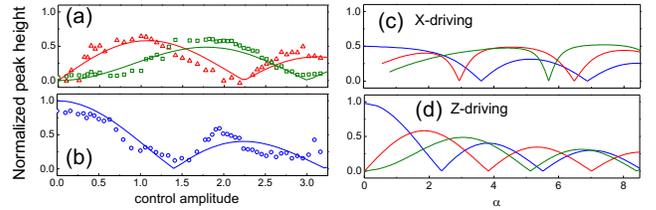}
\caption{
(a)(b) The peak height as a function of driving microwave amplitude for 1-photon(blue), 2-photon(red) and 3-photon(green) processes. The data sets are vertically and horizontally scaled by identical parameters. The solid curves are $|J_m|$ with $m=$ 0, 1 and 2. In (b) the dash curve shows the best fitting by using X-driving result. (c)(d) The calculated transmission amplitudes as a function of driving amplitude $\alpha$ for X-driving model (c) and Z-driving model (d).
}\label{fig3}
\end{figure}

The 2-tone spectroscopy measurements were performed with a low probe power and a driving microwave field at the frequency $\omega_d/2\pi=\omega_{2}/2\pi=$5.455 GHz. The probe frequency was chosen to be $\omega_{p}/2\pi=$5.513 GHz, about 1 MHz red detuned from $\omega_{1}/2\pi$. Such a detuning is small enough and allows us clearly to observe the transmon spectrum as it is shown as the black curve in Fig. \ref{fig2}(c). When the power of driving microwaves increases, the resonance peak associated with $|0\rangle$ to $|1\rangle$ transition becomes less pronounced. In addition, at the position where $|0\rangle$ to $|2\rangle$ transition appears in the 1-tone spectroscopy, a resonance structure starts to build up. However, these processes occur in an oscillatory way, rather than a monotonic way. Thus the resonance peaks associated with $|0\rangle$ to $|1\rangle$ and $|0\rangle$ to $|2\rangle$ transitions respectively vanish completely
and microwave transmission in these resonance channels are shut off at the powers of $P_d=$-102 dBm and -100 dBm.  
To investigate how the resonance structures are modulated by the driving power, we plot the peak heights as a function of the driving amplitude as illustrated in Fig. \ref{fig3}(a) and (b). The $|0\rangle$ to $|1\rangle$ channel reaches zero peak height first, then the  $|0\rangle$ to $|2\rangle$ channel and followed by $|0\rangle$ to $|3\rangle$ channel. The modulated resonance transmission of $|0\rangle$ to $|1\rangle$ transition and the occurrence of $|0\rangle$ to $|2\rangle$ and $|0\rangle$ to $|3\rangle$ channels suggest that the probe cavity strongly interacts with new quantum states, which are accounted as the dressed states due to the strong interaction of the bare transmon and driving microwaves.

Although several works have pointed out that the X-driving has an exact numerical solution, let's focus first on the Z-driving model, which provides an analytical description of the dressed states and allows a simple analysis for the major feature of the problem: 
$
|\pm,N\rangle=e^{\mp \eta\left(a^\dag-a\right)}|\pm\rangle|N\rangle.
$
$|\pm\rangle$ are the eigenstates of $\sigma_z$ associated with eigenvalues of $\pm1$.
When the dressed state is coupled to the probe cavity, the Hilbert space is expanded to include the degree of freedom originated from the probe photon number $n_p$. Because the probe microwave is weak, we may stick to the two following (probe) cavity photon number states, $n_p$ =0 and 1. The expanded quantum states are noted as, $|\pm, N, n_p=0\rangle$ and $|\pm, N, n_p=1\rangle$ with energies respectively $N\mp\varepsilon_0-g^2$ and $N\mp\varepsilon_0+1+\delta$. Here $\delta=(\omega_p-\omega_a)/\omega_d$ is the detuning of the probe microwave. The driving detuning $\Delta =(\omega_d-\omega_a)/\omega_d=1-2\lambda$. 

When the transmon is biased in such a way that its energy level spacing between ground state and $m$-th excited state obeying 
$2\varepsilon_0=\omega_{m0}/\omega_d\sim m$, two dressed states  $|+,N,1\rangle$ and $|-,N-m+1,0\rangle  $ become in resonance and are mixed with each other with a coupling strength, $g\times  \langle +,N-m+1| \sigma_x |-,N\rangle=gJ_{m-1}(\alpha)$ where $\alpha=4g\sqrt{N}$:
$$
\frac{H_{eff}}{\hbar\omega_d}=\frac{\omega_{0}}{\omega_d}+\left(
\begin{array}{cc}
\delta_m & gJ_{m-1}\\
gJ_{m-1} & (1-m)\Delta_m
\end{array}
\right).
$$
Here detunings $\Delta_m= 1-(\omega_{m0}/m\omega_{d})$ , $\delta_m= (\omega_p/\omega_{d})-(\omega_{m0}/m\omega_{d})$ and $\omega_0\sim N\omega_d+(\omega_{m0}/m)$. Because the driving detuning obeys $\Delta>\sim g>\delta$, we may diagonalize the matrix and obtain the lower energy state as
$
|\psi_0\rangle\sim  \left(|+, N, 1\rangle+|-, N-m+1, 0\rangle \right)/{\sqrt2}.
$
According to the linear response theory, the transmission of the probe microwave is given by the matrix element: 
$$
T\propto \langle \psi_0| \sigma_x|\psi_0 \rangle\sim\langle -,N |\sigma_x|+, N-m+1\rangle=J_{m-1}(4 \eta\sqrt{N}).
$$ 

In the case of X-driving,  the dressed state energy is 
$
E(\pm,N)/\hbar\omega_d=N\pm\lambda J_0(\alpha)+O(\eta^2)
$
with $\lambda=\omega_a/2\omega_d$. Regardless how large is the atom level spacing,
the two dressed states in resonance would be $|+, N, 1\rangle$ and $|+, N+1, 0\rangle$ with the coupling strength of 
$g\times \langle +, N+1|\sigma_x |+, N\rangle$. The lower mixed state is
$
|\psi_0\rangle\sim  \left(|+, N, 1\rangle+|+, N+1, 0\rangle \right)/{\sqrt2},
$
at which the transmission of the probe microwave is 
$$
T\propto \langle \psi_0| \sigma_x |\psi_0 \rangle= \langle +, N+1| \sigma_x|+, N\rangle.
$$ 
The results were calculated numerically when $\omega_a\sim m\omega_d$ and illustrated in Fig. 4(c).

To summarize, we should find that the transmission amplitude, $|T|$ of $|0\rangle$ to $|m\rangle$ channels which associated with $m$-photon dressed states should follow the amplitude ($\sqrt{\langle N\rangle}$) dependence of the Bessel functions $|J_{m-1}|$ if Z-driving is considered. Oscillatory behavior can also be seen for the X-driving case. Indeed, Fig. 4(a) and (b) show the high accordance of the theoretical prediction with the experimental data, which has a single scaling parameter in the driving amplitude. We also note that our scheme successfully demonstrated the Jaynes-Cumming model controlled by the driving tone whenever the probe tone is a fraction of the atom energy $\omega_p\sim\omega_a/m$.

%\section{conclusion}
In conclusion, we have studied the microwave dressed states of a transmon by coupled them with a cavity. By taking advantage of the spectrum structure, we may slightly tune the transmon in resonance to the $m$-photon process channels with a higher excited state $|m\rangle$. From the probe microwave transmission which is in resonance to the probe cavity, the structure of the dressed states is revealed. The transmission amplitudes through $m$-photon channels follow the Bessel function $|J_{m-1}|$ of the driving microwave amplitude. At the zeros of Bessel functions, the output probe microwaves are completely suppressed, leading to the controlled darkening of these resonance channels. The oscillatory behavior in transmission can be understood in the scope of LZS interferometry. The output of probe photons also follows  $m$-photon process and can find its application for a tunable source of correlated $m$ photons.

\begin{acknowledgments}
Fruitful discussions with M. C. Chung and K. Y. Chen are acknowledged. This work is financially supported by the Ministry of Science and Technology, Taiwan under grant No. 102-2628-M-005-001-MY4, and Research Center for Sustainable Energy and Nanotechnology, NCHU.
\end{acknowledgments}

\renewcommand\refname{Reference}

\bibliography{dress}

\end{document}